\documentclass[aps,nofootinbib,superscriptaddress,twocolumn]{revtex4}
\usepackage{graphicx,subfigure}
\usepackage{amssymb}
\usepackage{bm}
\usepackage{array}
\usepackage{slashed}
\usepackage{braket}
\usepackage{multirow}
\usepackage{bigstrut}
\usepackage[nooneline,footnotesize,center]{caption2}
\usepackage{longtable}
\usepackage{hyperref}
\usepackage{amsmath}

\begin{document}

\title{Quark-antiquark asymmetry of helicity distributions in the nucleon sea}

\author{Mengyun Liu}
\affiliation{School of Physics and State Key Laboratory of Nuclear Physics and Technology, Peking University, Beijing 100871, China}

\author{Bo-Qiang Ma}
\email{mabq@pku.edu.cn}
\affiliation{School of Physics and State Key Laboratory of Nuclear Physics and Technology, Peking University, Beijing 100871, China}
\affiliation{Collaborative Innovation Center of Quantum Matter, Beijing, China}
\affiliation{Center for High Energy Physics, Peking University, Beijing 100871, China}


\begin{abstract}
We study the helicity distributions of light flavor quark-antiquark ($q \bar{q}$) pairs in the nucleon sea. The valence quarks are handled by adopting the light-cone SU(6) quark-spectator-diquark model and the sea $q \bar{q}$ pairs are treated from statistical consideration by introducing
the helicity suppression factors $l_q(x)$ and $\bar l_q(x)$ to parametrize the helicity distributions of  q-flavor sea quark and antiquark respectively, while $\Delta l_q(x)=l_q(x)-\bar l_q(x)$ represents a combined effect of helicity contribution due to sea $q \bar{q}$ pairs. From fitting the nucleon polarization asymmetries $A^N_1$ in inclusive deep inelastic scattering processes and the single-spin asymmetries $A^{W^{\pm}}_L$ in Drell-Yan type processes, we find a significant asymmetry between the
quark and antiquark helicity distributions of the nucleon sea. Therefore the quark-antiquark asymmetry of helicity distributions of nucleon sea $q \bar{q}$ pairs, i.e.,
$\Delta q_s(x) \neq \Delta \bar q_s(x)$, plays an important role for a comprehensive understanding of the nucleon spin content.
\end{abstract}



\maketitle

\section{Introduction}
\label{introduction}

The spin structure of hadrons has received considerable attention since the so-called ``proton spin crisis"~\cite{Ashman:1987hv,Ashman:1989ig}, which implies that only a small fraction~(about $30\%$ in recent studies~\cite{Ageev:2005gh,Alexakhin:2006oza,Airapetian:2006vy}) of the proton spin comes from quark spins. Such a small quark spin contribution seems to contradict with the naive quark model~\cite{GellMann:1964nj,Zweig:1981pd} where all of the proton spin is provided by valence quark spins. However, it was pointed out in Refs.~\cite{Ma:1991xq,Ma:1992sj,Ma:1996np,Ma:1997gy} that the quark helicity observed in polarized deep inelastic scattering~(DIS) is actually the quark spin defined in light-form dynamics~\cite{Dirac:1949cp} and it is different from that in the quark model, which is defined in instant-form dynamics. Therefore the small helicity sum observed by the experiments is not in conflict with the quark model due to the reduction of the light-cone spin relative to the instant-form spin by the Melosh-Wigner rotation~\cite{Wigner:1939cj,Melosh:1974cu,Buccella:1974bz}, which is a relativistic effect of quark transversal motions. Calculations of the helicity distributions by valence quarks in the light-cone quark-spectator-diquark model~\cite{Ma:1996np,Chen:2005jsa} can reasonably reproduce the bulk features of the experimental data of spin asymmetries on  proton~\cite{Ashman:1987hv,Adeva:1993km,Adams:1994zd,Abe:1994cp}, neutron~\cite{Anthony:1993uf} and deuteron~\cite{Adeva:1993km,Adams:1994zd,Abe:1994cp} targets in inclusive DIS processes.

Nevertheless, the sea quark spin contribution to the proton is believed to be nontrivial, as reflected in the recent observations~\cite{Tian:2017xul,Tian:2017qwk} of single spin asymmetries of $W^{\pm}$ production in polarized proton-proton collisions~\cite{Adamczyk:2014xyw,Adamczyk:2015gyk}. It is shown~\cite{Tian:2017xul} that a reasonable description of $W^{\pm}$ single spin asymmetry data~\cite{Adamczyk:2014xyw} can be achieved by adopting sizable helicity distributions of the light-flavor up~(u) and down~(d) antiquarks. More explicitly, the antiquark helicity is positive for the $\bar{u}$ quark~($\Delta \bar u>0$) and negative for the $\bar{d}$ quark~($\Delta \bar d<0$). By adopting the quark-antiquark symmetry of the momentum and helicity distributions between quark-antiquark ($q \bar{q}$) pairs of the nucleon sea, it is found that the Melosh-Wigner rotation effect~\cite{Ma:1991xq,Ma:1992sj,Ma:1996np,Ma:1997gy} should be much stronger (with larger quark transversal motions) for the valence quarks to reconcile with both data from inclusive DIS processes~\cite{Adolph:2015saz} and $W^{\pm}$ productions~\cite{Adamczyk:2014xyw}.

From theoretical considerations, the quark and antiquark of the nucleon sea do not to be symmetric due to the nonperturbative nature of strong interaction. For example, the strange-antistrange asymmetry of the nucleon sea~\cite{Brodsky:1996hc,Signal:1987gz,Burkardt:1991di,Szczurek:1996pz} has been discussed to study the nucleon strange magnetic moment~\cite{Ma:1997gh},
the NuTeV anomaly~\cite{Ding:2004ht,Ding:2004dv,Wakamatsu:2004pd,Ding:2005ub,Alwall:2005xd,Wakamatsu:2014asa}, strange distribution functions of the nucleon~\cite{Barone:1999yv,Bourrely:2007if,Vega:2015hti}, nucleon form factors~\cite{Diehl:2007uc,Hobbs:2014lea}, the direct production of D-meson~\cite{Gao:2005gj,Hao:2005dw,Gao:2007ht,Gao:2008ch} and
the $\Lambda/\bar \Lambda$ polarization~\cite{Zhou:2009mx,Chi:2014xba,Du:2017nzy}.
However, the situation becomes subtle for the quark-antiquark asymmetry of the light-flavor sea $q \bar{q}$ pairs because of the existence of the valence u and d quarks in the nucleon, as one may always define $u_v(x)=u(x)-\bar{u}(x)$ and $d_v(x)=d(x)-\bar{d}(x)$ as the valence part of the total $u(x)=u_{v}(x)+u_s(x)$ and $d(x)=d_{v}(x)+d_s(x)$ quark contributions with the assumption of symmetric quark and antiquark momentum distributions: $u_s(x)=\bar{u}(x)$ and $d_s(x)=\bar{d}(x)$. The same situation also occurs for the helicity distributions of the light-flavor sea quarks. Inspired by the strange-antistrange asymmetry of the nucleon sea of previous studies, we now introduce a quark-antiquark asymmetry of helicity distributions of the nucleon sea while keeping a quark-antiquark symmetry of the momentum distributions for simplicity. We show that the quark-antiquark asymmetry can provide us an interesting scenario of the nucleon sea for a simultaneous description of both experimental data from DIS processes~\cite{Adams:1994zd,Abe:1994cp,Ashman:1987hv,Anthony:1993uf,Baum:1983ha,Abe:1997cx} and $W^{\pm}$ productions~\cite{Adamczyk:2014xyw}.

In this work, we investigate the quark-antiquark asymmetry of helicity distributions of the nucleon sea with the valence quarks handled by the light-cone SU(6) quark-spectator-diquark model~\cite{Ma:1996np,Ma:1996ii,Ma:1997gy} and the sea quarks treated by introducing new parameters from statistical consideration.
Numerical results are presented by fitting both the nucleon polarization asymmetries $A_1^N$ in polarized DIS processes and the single-spin asymmetries $A_L^{W^\pm}$ in Drell-Yan type processes.
Section \ref{model} is a brief introduction of the conventional calculation about quark distributions by the light-cone quark-spectator-diquark model, and in this sector we get parton distribution functions~(PDFs) of both valence and sea quarks as well as polarized PDFs of valence quarks. The polarized PDFs of sea quarks and antiquarks are given in Sec.~\ref{pdf} from statistical consideration by introducing new parameters: helicity suppression factors $l_q(x)$ and $\bar l_q(x)$ and combined helicity suppression factors $\Delta l_q(x)$.
The numerical calculations of the nucleon polarization asymmetries $A_1^N$ and the single-spin asymmetries $A^{W^{\pm}}_L$ are presented in Secs.~\ref{dis} and \ref{drellyan} respectively. We find a significant asymmetry between the
quark and antiquark helicity distributions of the nucleon sea. The results show that the quark-antiquark asymmetry of helicity distributions of nucleon sea $q \bar{q}$ pairs, i.e.,
$\Delta q_s(x) \neq \Delta \bar q_s(x)$, plays an important role for a comprehensive understanding of the nucleon spin content.
Finally, we provide a summary.
\section{Light-cone quark-spectator-diquark model}
\label{model}

The quark-diquark model~\cite{Field:1976ve,Close:1973xw,Carlitz:1975bg,Kaur:1977ce,Schaefer:1988xs} is an effective tool to describe the deep inelastic scattering of leptons on nucleons, with a picture that a single quark is struck by the incident lepton while the remaining part is treated as a quasiparticle of spectator diquark providing the quantum numbers and absorbing nonperturbative effects of all spectating particles. The light-cone SU(6) quark-spectator-diquark model~\cite{Ma:1996np, Ma:1996ii, Ma:1997gy} is actually an updated version
by taking into account the relativistic effect of quark transversal motions, with successful applications in investigating hadron structures by calculating relevant physical quantities, such as helicity~\cite{Ma:1996np,Chen:2005jsa} and transversity~\cite{Ma:1997gy} distributions, form factors~\cite{Ma:2002ir,Ma:2002xu,Liu:2014npa,Zhang:2016qqg},  transverse momentum dependent parton distributions~(TMDs)~\cite{Lu:2004au,She:2009jq,Lu:2012ez,Lu:2012gu,Liu:2015eqa}, generalized parton distributions~(GPDs), and Wigner distributions~\cite{Liu:2015eqa}. It is convenient to calculate PDFs and polarized PDFs of valence quarks, and especially it is pointed out in Refs.~\cite{Ma:1991xq,Ma:1992sj,Ma:1996np,Ma:1997gy} that the relativistic effect due to the Melosh-Wigner rotation~\cite{Wigner:1939cj,Melosh:1974cu,Buccella:1974bz} plays an important role for describing the spin-related quantities such as the helicity distributions.

As discussions in previous works~(see Ref.~\cite{Ma:1996np} for example), we can get PDFs and polarized PDFs of valence quarks of the proton from the model:
\begin{equation}\label{TheoryV}
	\begin{split}
		u_v(x) &= \frac{1}{2}a_S(x)+\frac{1}{6}a_V(x), \\
		d_v(x) &= \frac{1}{3}a_V(x), \\
		\Delta u_v(x) &= \left[u_v(x)-\frac{1}{2}d_v(x)\right]W_S(x)-\frac{1}{6}d_v(x)W_V(x), \\
		\Delta d_v(x) &= -\frac{1}{3}d_v(x)W_V(x),
	\end{split}
\end{equation}
where $x$ is the light-cone momentum fraction of the quark relative to the nucleon. $a_D(x)$~($D=S$ for scalar diquark and $D=V$ for vector diquark) is the probability when the quark $q$ is struck while the diquark state is $D$.
$a_D(x)$ can be written as
$
	a_D(x)\propto \int\left[\mathrm{d}^2\boldsymbol{k}_\perp\right]\left|\varphi_D(x, \boldsymbol{k}_\perp)\right|^2
$
with the normalization form $ \int^1_0a_D(x)\mathrm{d}x = 3$ and $\boldsymbol{k}_\perp$ represents the intrinsic transverse momentum of the quark.
$\varphi_D(x, \boldsymbol{k}_\perp)$ is the momentum space wave function which we adopt the Brodsky-Huang-Lepage~(BHL) prescription~\cite{Brodsky:1981jv,Huang:1994dy} for light-cone formalism: $\varphi_{D} (x, {\mathbf k}_\perp) = A_D \exp \left(-\mathcal{M}^2/8\beta_D^2\right)$, where $\beta_D$ is the harmonic oscillator scale parameter, $A_D$ is the normalization constant and $\mathcal{M}$ is the invariant mass: $\mathcal{M}^2 =  \left(m_q^2+\boldsymbol{k}_\perp^2\right)/x+\left(m_D^2+\boldsymbol{k}_\perp^2\right)/(1-x)$ where $m_q$ is the quark mass and $m_D$ is the diquark mass. $W_D(x, \boldsymbol{k}_\perp) = \left[(k^++m_q)^2-\boldsymbol{k}^2_\perp\right]/\left[(k^++m_q)^2+\boldsymbol{k}^2_\perp\right]$ is the correction factor due to the Melosh-Wigner rotation~\cite{Ma:1991xq,Ma:1992sj,Ma:1996np,Ma:1997gy} with $k^+=x\mathcal{M}$.  In this paper, we take the parameter values~(see Table~\ref{table:parameters}) the same as that given in Set 1 of Ref.~\cite{Zhang:2016qqg}, which studies the electromagnetic and weak form factors of the ground state octet
baryons using the light-cone quark-diquark model and gives a consistent description of the electroweak properties of the baryons in the low momentum transfer region.
\begin{table}
\renewcommand\arraystretch{1.2}
\caption{Parameters used in model calculations.}
\begin{center}
\begin{tabular}{c|cccccc}
\hline
\centering Quantity
&\centering{$m_q$}
&\centering $m_S$
&\centering $m_V$
&\centering $\beta_S$
&\centering $\beta_V$
&
\\ \hline
\centering Value~(MeV)
&\centering 330
&\centering 600
& 800
& 330
& 330
\\\hline
\end{tabular}
\end{center}
\label{table:parameters}
\end{table}

In order to get a balance between experimental data and model results, we adopt the following parametrization:
\begin{equation}
\begin{split}
\label{para}
u^{\text{para}}_v(x) &= u^{\text{CTEQ}}(x)- \bar u^{\text{CTEQ}}(x),\\
d^{\text{para}}_v(x) &= \frac{d^{\text{th}}_v(x)}{u^\text{th}_v(x)}u^{\text{para}}_v(x),\\
\Delta u^{\text{para}}_v(x) &= \left[u^{\text{para}}_v(x)-\frac{1}{2}d^{\text{para}}_v(x)\right]W_S(x)-\frac{1}{6}d^{\text{para}}_v(x)W_V(x), \\
\Delta d^{\text{para}}_v(x) &= -\frac{1}{3}d^{\text{para}}_v(x)W_V(x), \\
u^{\text{para}}_s(x) &= \bar u^{\text{para}}_s(x)  = \bar u^{\text{CTEQ}}(x),\\
d^{\text{para}}_s(x) &=\bar d^{\text{para}}_s(x) =  \bar d^{\text{CTEQ}}(x),\\
\end{split}
\end{equation}
where the superscript ``th'' means the pure theoretical results~[see Eq.~(\ref{TheoryV})] and ``CTEQ'' means CTEQ parametrization~\cite{Pumplin:2002vw}.

\section{Light-flavor sea (anti)quark helicity distributions}
\label{pdf}

In principle, there is no need to require ``the quark-antiquark
symmetry of the momentum distributions" for the light-flavor sea
quarks. There is no ambiguity for the strange-antistrange asymmetry
of the nucleon sea, as have been discussed in the literature.
However, the situation becomes complicated for the light-flavor u
and d cases as there are also valence u and d quarks inside the
nucleon, so it is hard to make a definite separation between the
sea part and the valence part for the total quark distribution
$q(x)=q_v(x)+q_s(x)$. One convenient definition is to assume ``the
quark-antiquark symmetry of the momentum distributions", i.e.,
$q_s(x)=\bar{q}(x)$ so that $q_v(x)=q(x)-\bar{q}(x)$. Other
definition with ``the quark-antiquark asymmetry of the momentum
distributions" is also possible with theory-dependent inputs, but
such asymmetry of the momentum distributions is small as from the
researches on the strange-antistrange asymmetry of the nucleon sea.
However, the quark-antiquark asymmetry might be significant in the
situation for the spin-dependent quantities. As a reasonable
approximation, we investigate the quark-antiquark asymmetry of
helicity distributions in the nucleon sea $q\bar q$ pairs while
keeping a quark-antiquark symmetry of the momentum distributions
as given in Eq.~(\ref{para}).

We assume that the sea helicity distributions are helicity-suppressed by the existence of same-flavor valence helicity distributions with the consideration of the Pauli blocking effect: the existence of a valence quark with certain polarization can cause a suppression of the same-flavor sea quark along the same polarization, and as the nonperturbative sea $q\bar{q}$ pairs are tend to have total spin zero configuration, so the antiquark tends to have the polarization parallel to the valence quark direction. Thus, we assume:
\begin{equation}\label{eq1}
  \begin{split}
    &\text{For sea quarks: }\\
      &\qquad q_v^\uparrow(x)\rightarrow l'_q(x)q_s^\downarrow(x)+\left[1-l'_q(x)\right]q_s^\uparrow(x), \\
      &\qquad q_v^\downarrow(x)\rightarrow l'_q(x)q_s^\uparrow(x)+\left[1-l'_q(x)\right]q_s^\downarrow(x); \\
    &\text{For sea antiquarks: }\\
      &\qquad q_v^\uparrow(x)\rightarrow \bar l'_q(x) \bar q_s^\uparrow(x)+\left[1-\bar l'_q(x)\right]\bar q_s^\downarrow(x),\\
      &\qquad q_v^\downarrow(x)\rightarrow \bar l'_q(x) \bar q_s^\downarrow(x)+\left[1-\bar l'_q(x)\right]\bar q_s^\uparrow(x),
	\end{split}
\end{equation}
where the right arrow$~(\rightarrow)$ indicates an impact of valence part distributions on sea part distributions, and $l'_q(x)$ is the probability of finding a sea quark $q_s$~($q=u, d$) with polarization antiparallel to the same-flavor valence quark $q_v$ at momentum fraction $x$, while $\bar l'_q(x)$ is the probability of finding a sea antiquark $\bar q_s$ with polarization parallel to the same-flavor valence quark. According to the Pauli blocking effect, we expect $0.5<l'_q(x)\leq 1$ and $0.5<\bar l'_q(x)\leq1$. Because $q(x) = q^\uparrow(x)+q^\downarrow(x)$ and $\Delta q(x) = q^\uparrow(x)-q^\downarrow(x)$, we can derive the following expressions with Eq.~$(\ref{eq1})$:
\begin{equation}\label{eq2}
\begin{split}
\Delta q_s(x) &\propto \frac{1}{1-2l'_q(x)}\frac{\Delta q_v(x)}{q_v(x)}q_s(x),\\
\Delta \bar q_s(x) & \propto -\frac{1}{1-2\bar l'_q(x)}\frac{\Delta q_v(x)}{q_v(x)}\bar q_s(x).
\end{split}
\end{equation}
For convenience, we assume:
\begin{equation}
\begin{split}
\Delta q_s(x) &= l_q(x)\frac{\Delta q_v(x)}{q_v(x)}q_s(x),\\
\Delta \bar q_s(x) &= -\bar l_q(x)\frac{\Delta q_v(x)}{q_v(x)}\bar q_s(x),
\end{split}
\end{equation}
where $l_q(x)$ and $ \bar l_q(x)$, which can be called helicity suppression factors, reflect the helicity suppression effects of q-flavor valence quark on the same-flavor sea quark and antiquark respectively. From Eq.~$(\ref{eq2})$ and the Pauli blocking effect, we expect $l_q(x)\leq0$ and $\bar l_q(x)\leq0$, which are confirmed by following numerical calculations~(see Table~\ref{table:result2}). It is obvious that $\left| l_q(x)\frac{\Delta q_v(x)}{q_v(x)}\right|\leq1$ and $\left| \bar l_q(x)\frac{\Delta q_v(x)}{q_v(x)}\right|\leq1$.

\begin{figure}[htbp]
\centering
\includegraphics[scale=0.35]{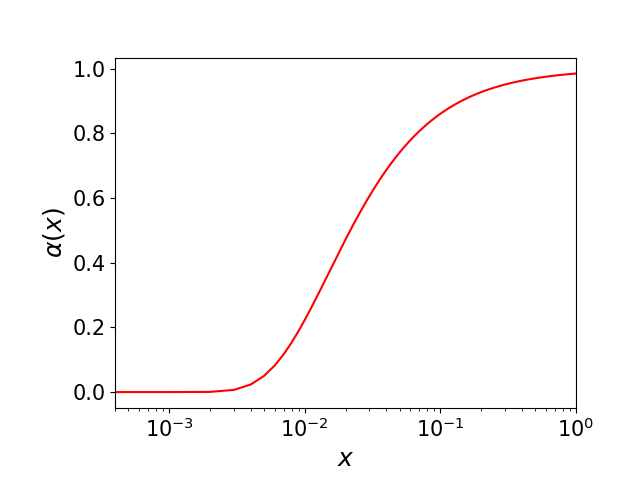}
\caption{The adjusting function $\alpha(x)$ as a function of the momentum fraction $x$. The function tends to 1 when $x$ is not so small, and to 0 when $x$ is very small.}
\label{fig:wcurve}
\end{figure}

Two schemes are used to solve $l_q(x)$ and $ \bar l_q(x)$:
\begin{enumerate}
\item
$\left\{
	\begin{array}{lr}
             l_q(x) &= l_q,  \\
             \bar l_q(x) &= \bar l_q,\\
             \end{array}
\right.
$for convenience.
\item
$\left\{
	\begin{array}{lr}
             l_q(x) = l_q\times \alpha(x),  \\
             \bar l_q(x) = \bar l_q\times \alpha(x),\\
             \end{array}
\right.
$
where $\alpha(x)$ is an adjusting function with the characteristic features: the function tends to 1 when $x$ is not so small, and to 0 when $x$ is very small. We recommend $\alpha(x) =\exp(-c x^{-1})$~(see Fig.~\ref{fig:wcurve}) with $c=0.015$ (which is adjustable) as an option. The reason for this choice is based on the consideration that the impact of valence part on sea part should be big when $x$ is not so small and little when $x$ is very small.
\end{enumerate}

Furthermore, because the quark-antiquark symmetry of the momentum distributions~[i.e., $q_s(x) = \bar q_s(x)$] is assumed, we can define:
\begin{equation}\label{SeaTotalDelta}
\begin{split}
\Delta q_{s+\bar s} &= \Delta q_s(x) + \Delta \bar q_s(x)= \Delta l_q(x)\frac{\Delta q_v(x)}{q_v(x)}q_s(x),\\
&\text{with}\quad\Delta l_q(x) = l_q(x)-\bar l_q(x),
\end{split}
\end{equation}
where $\Delta l_q(x)$, which can be called combined helicity suppression factor of quark $q$, represents the combined helicity suppression effect.

\section{Nucleon polarization asymmetries $A^N_1$ in polarized DIS processes}
\label{dis}
With the PDFs and polarized PDFs derived above and due to isospin symmetry between proton and neutron, we can get spin-independent structure functions $F^N_2(x)$~($N=p,n$):
\begin{equation}
\begin{split}
F^p_2(x) &= x\left\{\frac{4}{9}\left[u(x)+\bar u(x)\right] + \frac{1}{9}\left[d(x)+\bar d(x)\right]\right\}, \\
F^n_2(x) &= x\left\{\frac{1}{9}\left[u(x)+\bar u(x)\right] + \frac{4}{9}\left[d(x)+\bar d(x)\right]\right\}, \\
\end{split}
\end{equation}
where $q(x) = q_v(x)+q_s(x), ~\bar q(x) = \bar q_s(x)$ and spin-dependent structure functions  $g^N_1(x)$~($N=p,n$):
\begin{equation}
\begin{split}
&g_1^p(x) = \frac{1}{2}\\
&\times \left\{\frac{4}{9}\left[\Delta u_v(x)+\Delta u_{s+\bar s}(x)\right]+\frac{1}{9}\left[\Delta d_v(x)+\Delta d_{s+\bar s}(x)\right]\right\},\\
&g_1^n(x)  =\frac{1}{2} \\
&\times\left\{\frac{1}{9}\left[\Delta u_v(x)+\Delta u_{s+\bar s}(x)\right]+\frac{4}{9}\left[\Delta d_v(x)+\Delta d_{s+\bar s}(x)\right]\right\},
\end{split}
\end{equation}
where $\Delta q_{s+\bar s}$ is defined in Eq.~$(\ref{SeaTotalDelta})$. The nucleon polarization asymmetries $A^N_1$~($N =p,n$) are directly measured in experiments and expressed as
$
A^N_1(x) = 2xg^N_1(x)/F^N_2(x)
$.

In the nucleon sea, the polarization asymmetries $A^N_1$ are only sensitive to the total helicity distribution functions~($\Delta q_{s+\bar s}$) so that we only need to fit the combined helicity suppression factors $\Delta l_q(x)$~($q=u, d$).
We get $\Delta l_q(x)$~($q=u, d$) by fitting experimental data of proton polarization asymmetry $A^p_1$ at E130~\cite{Baum:1983ha}, EMC~\cite{Ashman:1987hv}, SMC~\cite{Adams:1994zd}, E143~\cite{Abe:1994cp} and neutron polarization asymmetry $A^n_1$ at E142~\cite{Anthony:1993uf}, E154~\cite{Abe:1997cx}. The fitting method is:
\begin{align}
\label{fittingmethod}
\chi^2&=\sum_{i}(\mathcal{O}_i^{\text{exp}}
-\mathcal{O}_i^{\text{th}})^2({\sigma_i^2})^{-1},
\end{align}
where $\mathcal{O}_i^{\text{exp}}$ and $\mathcal{O}_i^{\text{th}}$ are experimental and theoretical values at the $i$'s data respectively, and $\sigma_i$ is $i$'s experimental uncertainty.

\begin{figure*}[htbp]\centering
\subfigure[Scheme 1: proton spin asymmetry $A^p_1(x)$.]{\begin{minipage}{7cm}\centering\includegraphics[scale=0.34]{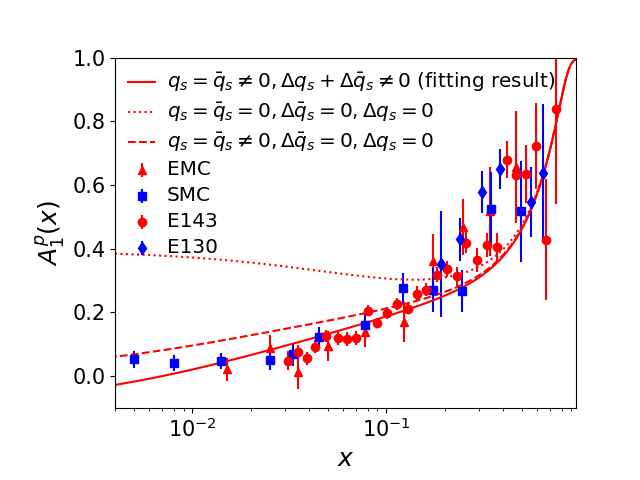}\end{minipage}}
\subfigure[Scheme 2: proton spin asymmetry $A^p_1(x)$.]{\begin{minipage}{7cm}\centering\includegraphics[scale=0.34]{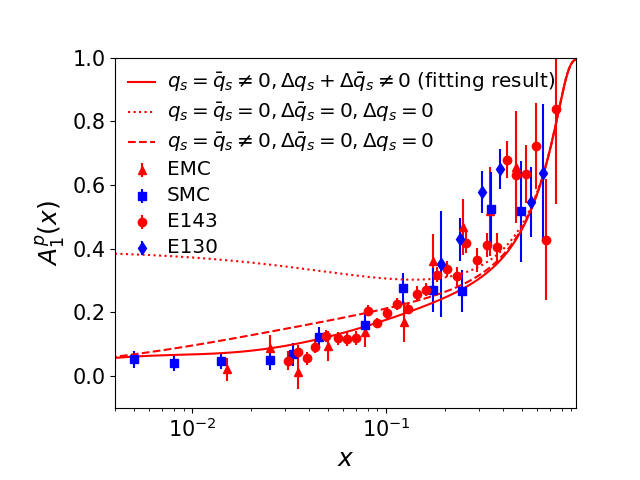}\end{minipage}}

\subfigure[Scheme 1: neutron spin asymmetry $A^n_1(x)$.]{\begin{minipage}{7cm}\centering\includegraphics[scale=0.34]{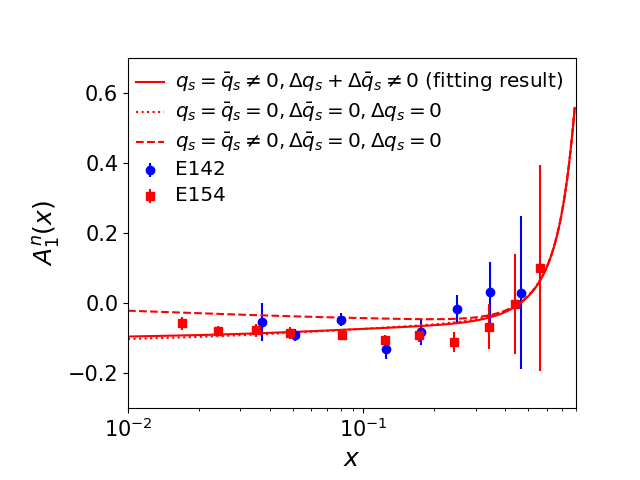}\end{minipage}}
\subfigure[ Scheme 2: neutron spin asymmetry $A^n_1(x)$.]{\begin{minipage}{7cm}\centering\includegraphics[scale=0.34]{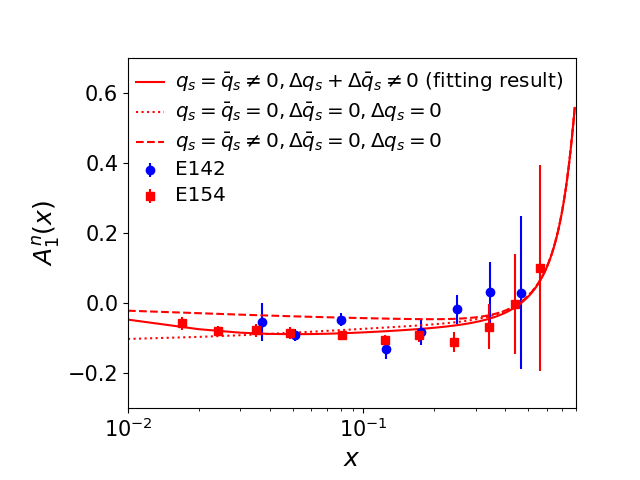}\end{minipage}}
\caption{The nucleon polarization asymmetries $A^N_1(x)$: (a) and (b) for the proton with data from SMC~\cite{Adams:1994zd}, E143~\cite{Abe:1994cp}, E130~\cite{Baum:1983ha}, EMC~\cite{Ashman:1987hv}; (c) and (d) for the neutron with data from E142~\cite{Anthony:1993uf}, E154~\cite{Abe:1997cx}.
The dotted curves are the results without considering the sea contributions. The dashed curves take into account the contributions of unpolarized distributions of sea part but not the polarized ones. The solid curves consider both unpolarized and polarized sea contributions and are fitting results.}
\label{fig:a1N}
\end{figure*}

The fitting results of combined helicity suppression factors~($\Delta l_u$ and $\Delta l_d$) are exhibited in Table~\ref{table:result1}. Because $\Delta u_v>0$ and $\Delta d_v<0$, we come to the conclusion that $\Delta u_{s+\bar s}<0$ and $\Delta d_{s+\bar s}<0$. The calculated nucleon polarization asymmetries of proton~$A^p_1$ and neutron~$A^n_1$ as functions of $x$ are shown in Fig.~\ref{fig:a1N} and compared with experimental data. By considering sea helicity distributions, the theoretical calculation results of $A_1^N$ are better consistent with the experimental data. Two schemes are adopted and a noticeable deviation appears at small $x$. The fitting results~(solid curves) of Scheme 2 are in good agreement with the experimental data, while the results of Scheme 1 deviate from the experimental data at small $x$.
\begin{table}
\renewcommand\arraystretch{1.2}
\caption{Fitting results of combined helicity suppression factors~($\Delta l_u$ and $\Delta l_d$).}
\begin{center}
\begin{tabular}{p{2cm}p{2cm}p{2cm}c}
\hline
\centering SCHEME
&\centering{$\Delta l_u$}
&\centering $\Delta l_d$
&
\\ \hline
\centering $1$
&\centering -0.421
&\centering 0.586
&
\\
\centering 2
&\centering -0.713
&\centering 0.885
&
\\\hline
\end{tabular}
\end{center}
\label{table:result1}
\end{table}

In addition, because the spin dependent structure functions $g_1^N(x)$ are related to the net quark helicity in nucleon, the first moments $\Gamma^N_1 = \int_0^{1}g_1^N(x)\mathrm{d}x$ are checked. In Table~\ref{thetable}, we give the calculation results  without the contribution of the sea part and the results with the contribution of the sea part of two schemes.   By comparing with results of COMPASS~\cite{Adolph:2015saz}, we notice that the theoretical results of Scheme 1 are in agreement with experimental results when $x$ is not so small~[$x>0.0025(0.004)$], but deviate from experimental results when $x$ is very small~[$x<0.0025(0.004)$]. This kind of behavior is consistent with that analyzed above when discussing the nucleon polarization asymmetries. The behavior of Scheme 1 may be caused by the changeless influence of valence part on sea part and this hypothesis of the changeless influence is perhaps inadequate because of the considerable parton distribution of sea quarks compared to the valence part at small $x$. The situation is improved when it comes to Scheme 2, which is exactly the one that is arranged with a variable valence influence. We notice that the results of Scheme 2 are in good agreement with COMPASS results~\cite{Adolph:2015saz} in each interval of $x$ .

\begin{table*}
\renewcommand\arraystretch{1.1}
\caption{The results of $\Gamma^p_1$, $\Gamma^n_1$ and $\Gamma^N_1$ ($\Gamma^N_1 = \frac{1}{2}(\Gamma^p_1+\Gamma^n_1)$).}
\begin{center}
\begin{tabular}{c|l|c|c|c|c}
\hline
& $x$ range
& Scheme 1
& Scheme 2
& Without sea part
& COMPASS\cite{Adolph:2015saz}
\\\hline
\multirow{4}{*}{$\Gamma^p_1$}
&0-0.0025
& -0.031
& 0.012
& 0.012
& 0.002
\\
&0.0025-0.7
&0.115
&0.134
&0.159
& $0.134\pm 0.003$
\\
&0.7-1.0
&0.001
&0.001
&0.001
&0.003
\\
&0-1
&0.086
&0.147
&0.172
&$0.139\pm0.006$
\\\hline
\multirow{4}{*}{$\Gamma^N_1$}
&0-0.004
&-0.042
& 0.006
&0.006
&0.000
\\
& 0.004-0.7
& 0.029
& 0.041
&0.065
&$0.047\pm0.003$
\\
& 0.7-1
&0.001
&0.001
&0.001
&0.001
\\
& 0-1
& -0.012
& 0.048
& 0.073
& $0.049\pm0.003$
\\\hline
$\Gamma^n_1$
& 0-1
& -0.110
& -0.052
&-0.027
&$-0.041\pm 0.006$
\\\hline
\end{tabular}
\end{center}
\label{thetable}
\end{table*}

We modify the nucleon spin asymmetries by considering the helicity distributions of sea quarks and antiquarks. Nevertheless, we only get the combined helicity suppression factors~[$\Delta l_u(x)$ and $\Delta l_d(x)$], but having no idea about the single helicity suppression effects of sea quarks as well as antiquarks. We will check these effects in the following section by calculating single-spin asymmetries $A^{W^{\pm}}_L$ in Drell-Yan type processes.

\section{Single-spin asymmetries $A^{W^{\pm}}_L$ in $\vec{p}+p$ collisions }
\label{drellyan}
Due to the pure V-A structure of weak interaction vertex $Wq\bar q'$ and because of quark helicity conservation at the  vertex,  the single-spin asymmetries  $A^{W^+}_L$ is sensitive to  $\Delta \bar d_s(x)$ and $\Delta u_s(x)$, while $A^{W^-}_L$ to $\Delta \bar u_s(x)$ and $\Delta d_s(x)$, which allow us to explore the helicity suppression effects of sea quarks and antiquarks individually.

\begin{figure*}[htbp]\centering
\subfigure[Scheme 1: single-spin asymmetries $A^{W^{+}}_L$.]{\begin{minipage}{7.2cm}\centering\includegraphics[scale=0.35]{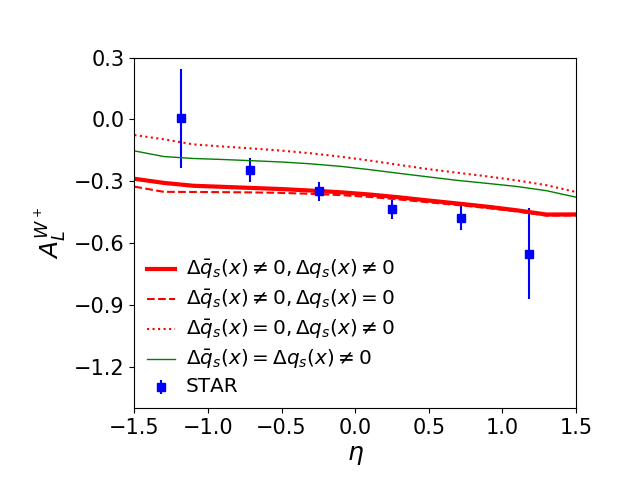}\end{minipage}}
\subfigure[ Scheme 2: single-spin asymmetries $A^{W^{+}}_L$.]{\begin{minipage}{7cm}\centering\includegraphics[scale=0.35]{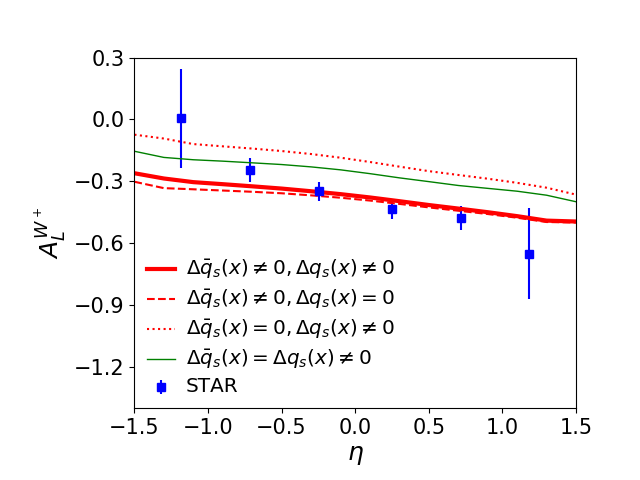}\end{minipage}}

\subfigure[ Scheme 1: single-spin asymmetries $A^{W^{-}}_L$.]{\begin{minipage}{7.2cm}\centering\includegraphics[scale=0.35]{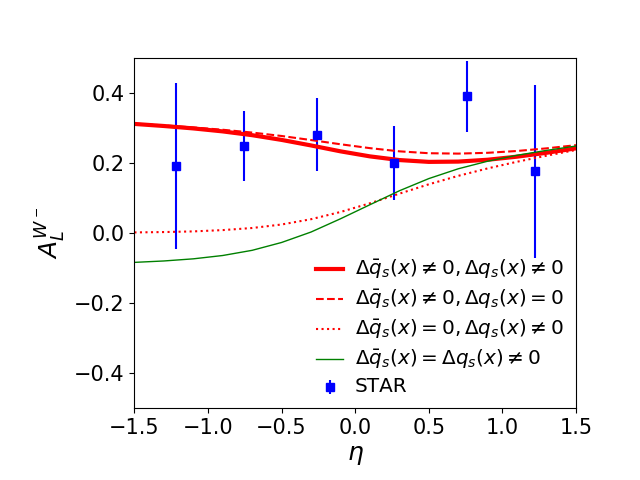}\end{minipage}}
\subfigure[ Scheme 2: single-spin asymmetries $A^{W^{-}}_L$.]{\begin{minipage}{7cm}\centering\includegraphics[scale=0.35]{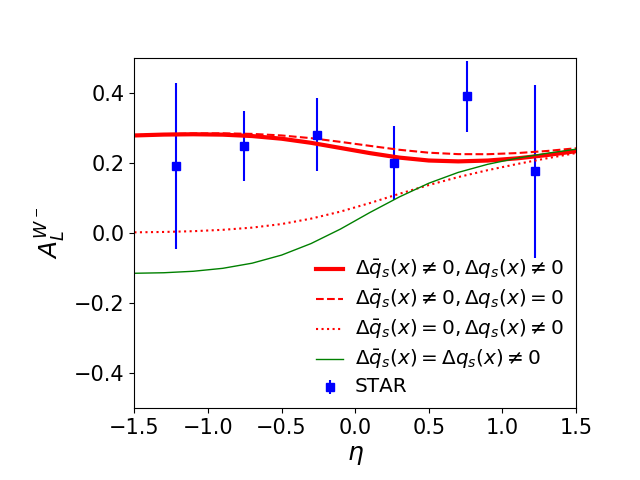}\end{minipage}}
\caption{The single-spin asymmetries $A^{W^{\pm}}_L$ as a function of the lepton pseudorapidity $\eta$. The data are from \cite{Adamczyk:2014xyw}. The thick (red) solid curves are the fitting results, while the thin (green) solid curves adopt the assumption $\Delta q_s(x) = \Delta \bar q_s(x)$ obtained by $\Delta l(x)$. Dotted curves only consider the sea quark helicity distributions, and dashed curves only consider the sea antiquark helicity distributions.}
\label{fig:awpm}
\end{figure*}

At leading order~(LO), the single-spin asymmetry $A^{W^+}_L$ with midprocess $u\bar d \rightarrow W^+$ and $A^{W^-}_L$ with midprocess $d\bar u \rightarrow W^-$ can be expressed as:
\begin{equation}\label{aw1}
\begin{split}
A^{W^+}_L &= \frac{-\Delta u(x_1)\bar d(x_2) + \Delta \bar d(x_1)u(x_2)}{u(x_1)\bar d(x_2)+\bar d(x_1)u(x_2)},\\
A^{W^-}_L &= \frac{-\Delta d(x_1)\bar u(x_2) + \Delta \bar u(x_1)d(x_2)}{d(x_1)\bar u(x_2)+\bar u(x_1)d(x_2)},\\
\end{split}
\end{equation}
with
\begin{equation}\label{aw2}
\begin{split}
q(x) = q_v(x)+q_s(x),& ~~\bar q(x)  = \bar q_s(x), \\
\Delta q(x)=\Delta q_v(x)+\Delta q_s(x),& ~~\Delta \bar q(x) = \Delta \bar q_s(x),
\end{split}
\end{equation}
where $q=u,d$.
\newline\indent
With the combination of Eq.~$(\ref{SeaTotalDelta})$ and fitting results of $\Delta l_u(x)$ and $\Delta l_d(x)$~(see Table~\ref{table:result1}) together, a simplification can be made by fitting two parameters [$\bar l_u(x)$ and $\bar l_d(x)$] instead of four with the method declared in Eq.~$(\ref{fittingmethod})$. We get $\bar l_u$ and $ \bar l_d$ by fitting experimental data of single-spin asymmetries $A^{W^{\pm}}_L$ at RHIC~\cite{Adamczyk:2014xyw}.
The fitting results of $\bar l_u$ and $\bar l_d$ and calculated results of $l_u$ and $l_d$ are exhibited in Table~\ref{table:result2}. The calculated single-spin asymmetries~$A^{W^{\pm}}_L$ as functions of $x$ are shown in Fig.~\ref{fig:awpm} and compared with experimental data. More calculation details can be found in Refs.~\cite{Tian:2017xul,Ringer:2015oaa}.
\begin{table*}
\renewcommand\arraystretch{1.2}
\caption{Fitting results of $\bar l_u$ and $\bar l_d$ and calculated results of $l_u$ and $l_d$.}
\begin{center}
\begin{tabular}{p{2cm}p{2cm}p{2cm}p{2cm}p{2cm}c}
\hline
\centering SCHEME
&\centering{$\bar l_u$}
&\centering $\bar l_d$
&\centering $l_u$
&\centering $l_d$
&
\\ \hline
\centering $1$
&\centering -0.751
&\centering -1.347
&\centering -1.172
&\centering -0.761
&
\\
\centering 2
&\centering-0.836
&\centering-1.658
&\centering-1.549
&\centering-0.773
&
\\\hline
\end{tabular}
\end{center}
\label{table:result2}
\end{table*}

In Fig.~\ref{fig:awpm}, the curves are calculated with the results of combined suppression factors~[$\Delta l_u(x)$ and $\Delta l_d(x)$] obtained in Sec.~\ref{dis}. The thin-solid~(green) curves are directly calculated by adopting a quark-antiquark symmetry of helicity distributions~[i.e., $\Delta q_s(x) = \Delta \bar q_s(x)$], while the thick-solid, dotted and dashed~(red) curves are calculated by a second fitting without a preassumption of the quark-antiquark symmetry of helicity distributions, and the fitting results show a bulk of quark-antiquark asymmetry~[i.e., $\Delta q_s(x) \neq \Delta \bar q_s(x)$]. The bad performances of the thin-solid~(green) curves accord with the results in Ref.~\cite{Tian:2017xul}, which show that it is difficult to give a simultaneous description of both the experimental data from DIS processes and that from $W^\pm$ productions with a quark-antiquark symmetry of helicity distributions.  However, after taking into account the the quark-antiquark asymmetry of helicity distribution of nucleon sea $q\bar q$ pairs, the thick-solid~(red) curves are consistent with the experimental data. The thick-solid curves consider the helicity contributions of both the quarks and antiquarks in nucleon sea, while the  dotted curves and dashed curves only consider the helicity contributions of sea quarks and antiquarks respectively. The curves show that the contribution of helicity distributions of sea antiquarks are much bigger than that of sea quarks. Two different schemes make little difference and the Scheme 2 is a little better than Scheme 1. From the results shown in Table~\ref{table:result2}, the helicity suppression factors~($\bar l_u$, $\bar l_d$, $l_u$ and $l_d$) are all negative, which is expected as discussed in Sec.~\ref{pdf} from statistical consideration of the Pauli blocking effect.

\begin{figure*}[htbp]\centering
\setlength{\abovecaptionskip}{0.cm}
\setlength{\belowcaptionskip}{-0.cm}
\subfigure[ Scheme 1: $\frac{\Delta q(x)}{q(x)}$.]{\begin{minipage}{7.1cm}\centering\includegraphics[scale=0.34]{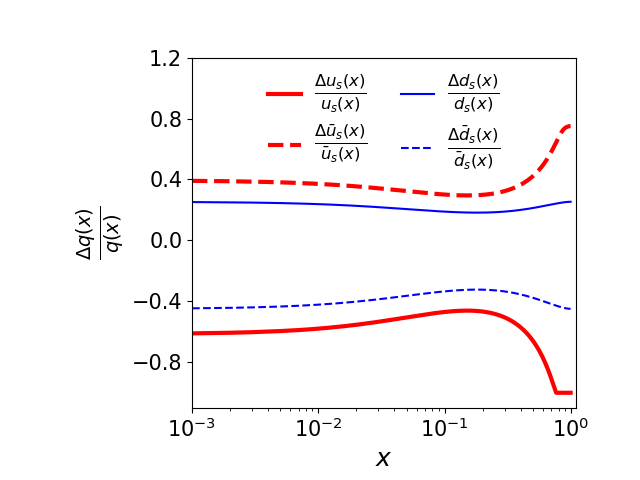}\end{minipage}}
\subfigure[ Scheme 2: $\frac{\Delta q(x)}{q(x)}$.]{\begin{minipage}{7cm}\centering\includegraphics[scale=0.35]{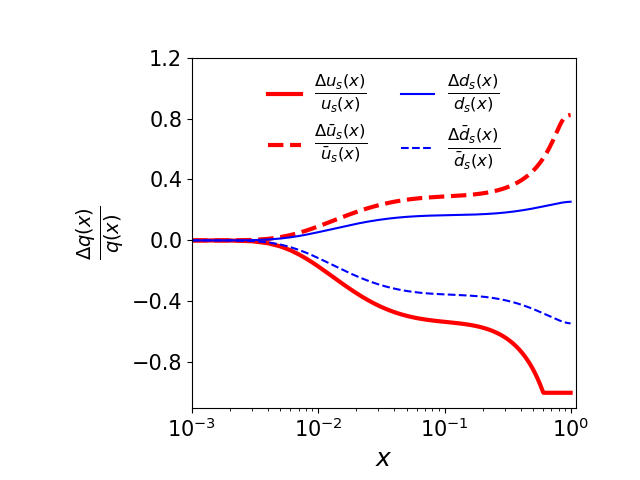}\end{minipage}}
\caption{$\frac{\Delta q(x)}{q(x)}$ as functions of the momentum fraction $x$ in two schemes.}
\label{fig:n}
\end{figure*}

\begin{figure*}[htbp]\centering
\subfigure[ Scheme 1: $x\Delta q(x)$.]{\begin{minipage}{7.1cm}\centering\includegraphics[scale=0.34]{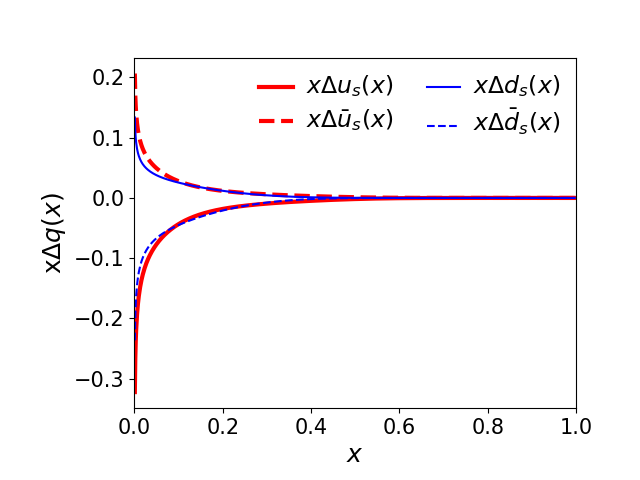}\end{minipage}}
\subfigure[ Scheme 2: $x\Delta q(x)$.]{\begin{minipage}{7cm}\centering\includegraphics[scale=0.35]{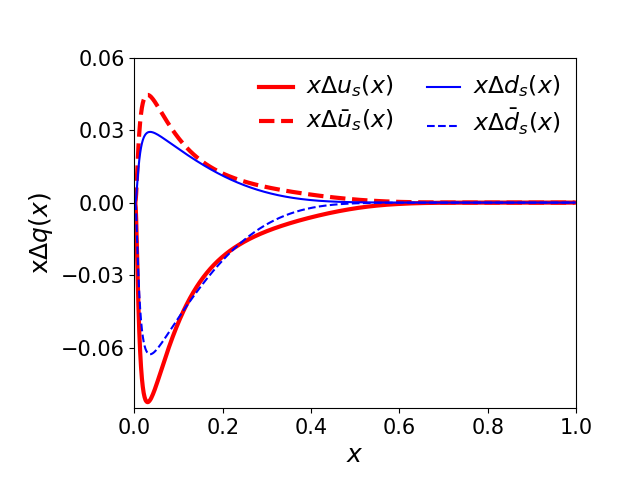}\end{minipage}}
\caption{$x\Delta q(x)$ as functions of the momentum fraction $x$ in two schemes.}
\label{fig:q}
\end{figure*}

Figure~$\ref{fig:n}$ shows the results of ${\Delta q(x)}/{q(x)}$ and  ${\Delta \bar q(x)}/{\bar q(x)}$ as functions of the momentum fraction $x$. Figure~$\ref{fig:q}$ shows the results of $x\Delta q(x)$ and $x\Delta \bar q(x)$ as functions of the momentum fraction $x$.
 The behaviors about $x\Delta q(x)$ and $x\Delta \bar q(x)$ are consistent with the statistical consideration~\cite{Bourrely:2001du,Bhalerao:2001rn,Gluck:2000ch}, the phenomenological analysis~\cite{Tian:2017xul} as well as the  parametrized results in Refs.~\cite{deFlorian:2009vb,Arbabifar:2013tma}, which study the asymmetries $A_1^N$, $A_1^{p,\pi^\pm}$ and $A_1^{p,K^\pm}$ in inclusive and semi-inclusive polarized DIS processes. The results also indicate a significant flavor asymmetry of the antiquark helicity distributions as predicted from some theoretical considerations~\cite{Diakonov:1996sr,Diakonov:1997vc,Wakamatsu:1999vf,Dressler:1999zg,Fries:2002um}.

The model calculation results of quantities $\Delta q$~($\Delta q = \int^1_0q(x)\mathrm{d}x$) are shown in Table~$\ref{table:result_q}$. The results of Scheme 1 and Scheme 2 are very different, especially when calculating $\Delta d$ and $\Delta\Sigma$~(for example, the result of  $\Delta\Sigma$ of Scheme 2 is positive, while that of Scheme 1 is negative). The difference of the results comes from the different impact of valence part on sea part of these two schemes as elaborated in Sec.~\ref{pdf}. In Scheme 1, the impact of valence part helicity distributions on sea part helicity distributions is changeless in the range of the entire independent variable $x$, while in Scheme 2, the impact is variable with $x$. Compared with parametrized results~($\Delta u^+$, $\Delta d^+$ and $\Delta \Sigma$) of NNPDFpl11.1~\cite{Nocera:2014gqa} and DSSV08~\cite{Hirai:2008aj}, Scheme 2 performs much better than Scheme 1 as discussed in Sec.~\ref{dis}, and the results of Scheme 2 are in pretty good agreement with these two parametrized results.

Back to the ``proton spin crisis", recent studies~\cite{Ageev:2005gh,Alexakhin:2006oza,Airapetian:2006vy} reported about $30\%$ of the proton spin coming from quark spins.
With the discussion about spin and dynamics given in Introduction, we can understand the ``spin crisis" by recognizing the quantity $\Delta \Sigma$~($\Delta \Sigma = \Delta u^+ +\Delta d^+$ with $\Delta q^+ = \Delta q_v+\Delta q_s+\Delta \bar q_s$) as the sum of quark helicities in light-form dynamics
rather than the vector sum of spins carried by quarks and antiquarks in the proton rest frame.  Therefore, the quantity $\Delta \Sigma$, representing the light-cone spin sum of quarks and antiquarks inside the proton, plays an important role in studying the nucleon spin structure. As shown in Table~$\ref{table:result_q}$, the theoretical calculation results of $\Delta \Sigma$ of Scheme 2 is 0.302, which just falls on the parametrized results: $0.25\pm0.10$ from NNPDFpl11.1~\cite{Nocera:2014gqa} and $0.366^{+0.042}_{-0.062}(+0.123)$ from DSSV08~\cite{Hirai:2008aj}. Thus Scheme 2 provides a reasonable scenario to calculate sea quark and antiquark helicity distributions from statistical consideration. The results indicate that the light-cone SU(6) quark-spectator-diquark model, together with
the relativistic effect due to quark transversal motions
and the quark-antiquark asymmetry of helicity distributions of the
nucleon sea with phenomenologically based models and parameters, can provide good descriptions for the physical quantities related to the ``proton spin crisis".

\begin{table*}[!htbp]
\renewcommand\arraystretch{1.2}
\caption{$\Delta q$, quantities from model calculations.}
\begin{center}
\begin{tabular}{p{3cm}p{0.5cm}p{0.5cm}p{0.5cm}p{0.5cm}p{0.5cm}p{0.5cm}p{0.5cm}p{0.5cm}p{0.5cm}p{0.5cm}p{0.5cm}p{0.5cm}}
 \rule{12.1cm}{0.05cm}
&&&&&&&\\
\setlength{\tabcolsep}{0pt}
\centering SCHEME
&\multicolumn{3}{c}{$\Delta u_v$}
&\multicolumn{3}{c}{$\Delta u_s$}
&\multicolumn{3}{c}{$\Delta \bar u_s$}
&\multicolumn{3}{c}{$\Delta u$}
\\ \hline
\centering $1$
&\multicolumn{3}{c}{ 0.827}
&\multicolumn{3}{c}{ -0.808}
&\multicolumn{3}{c}{ 0.518}
&\multicolumn{3}{c}{ 0.019}
\\
\centering 2
&\multicolumn{3}{c}{ 0.827}
&\multicolumn{3}{c}{-0.221}
&\multicolumn{3}{c}{ 0.120}
& \multicolumn{3}{c}{ 0.606}

\\\hline\hline
\centering SCHEME
&\multicolumn{3}{c}{$\Delta d_v$}
&\multicolumn{3}{c}{ $\Delta d_s$}
&\multicolumn{3}{c}{ $\Delta \bar d_s$}
& \multicolumn{3}{c}{$\Delta d$}
\\ \hline
\centering $1$
&\multicolumn{3}{c}{ -0.329}
&\multicolumn{3}{c}{ 0.353}
&\multicolumn{3}{c}{ -0.624}
& \multicolumn{3}{c}{ 0.024}
\\
\centering 2
&\multicolumn{3}{c}{ -0.329}
&\multicolumn{3}{c}{ 0.083}
&\multicolumn{3}{c}{ -0.177}
&\multicolumn{3}{c}{ -0.247}
\\\hline\hline
\centering SCHEME
&\multicolumn{4}{c}{$\Delta u^+$}&\multicolumn{4}{c}{$\Delta d^+$}&\multicolumn{4}{c}{$\Delta \Sigma$}
\\ \hline
\centering $1$
&\multicolumn{4}{c}{0.537}&\multicolumn{4}{c}{-0.601}&\multicolumn{4}{c}{-0.064}
\\
\centering 2
&\multicolumn{4}{c}{0.725}&\multicolumn{4}{c}{-0.424}
&\multicolumn{4}{c}{0.302}
\\
\centering NNPDFpl11.1~\cite{Nocera:2014gqa}
&\multicolumn{4}{c}{$0.76\pm0.04$}
&\multicolumn{4}{c}{$-0.41\pm0.04$}
&\multicolumn{4}{c}{$0.25\pm0.10$}
\\
\centering DSSV08~\cite{Hirai:2008aj}
&\multicolumn{4}{c}{$0.793^{+0.028}_{-0.034}(+0.020)$}
&\multicolumn{4}{c}{$-0.416^{+0.035}_{-0.025}(-0.042)$}
&\multicolumn{4}{c}{$0.366^{+0.042}_{-0.062}(+0.123)$}
\\\hline
\end{tabular}
\end{center}
\label{table:result_q}
\end{table*}
\section{Summary}
In summary, we study the helicity distributions of light-flavor $q\bar q$ pairs in the nucleon sea. We investigate the contributions of sea quark and antiquark helicity distributions to nucleon polarization asymmetries $A_1^N$ in DIS processes and single-spin asymmetries $A^{W^\pm}_L$ in Drell-Yan type processes. The results show that there is a significant asymmetry between quark and antiquark helicity distributions of the nucleon sea. We compare the model results of $\Gamma^N_1$, $x\Delta q(x)$ and $\Delta q$~($N=p,n$) with experimental~\cite{Adolph:2015saz}, parametrized~\cite{deFlorian:2009vb,Arbabifar:2013tma,Nocera:2014gqa,Hirai:2008aj}, statistical~\cite{Bourrely:2001du,Bhalerao:2001rn,Gluck:2000ch} and phenomenological~\cite{Tian:2017xul} results, and they conform well. From statistical consideration, we assume that the sea helicity distributions are helicity-suppressed by the existence of same-flavor valence helicity distributions. With this statistical assumption, we give the method of calculating helicity distributions of sea quarks and antiquarks by introducing new parameters: helicity suppression factors $l_q(x)$ and $\bar l_q(x)$ and combined helicity suppression factors $\Delta l_q(x)$. Two schemes are put forward to help calculations. The good performance of the second scheme show that because of the relative quantitative relation between valence quarks and sea (anti)quarks, the influence of valence part helicity distributions on sea part helicity distributions is variable with the momentum fraction $x$: the influence is small when $x$ is very small, while big when $x$ is not so small.
It is shown that the quark-antiquark asymmetry of helicity distributions of the nucleon sea is important for a simultaneous description of the nucleon polarization asymmetries $A^N_1$ in inclusive DIS processes and the single-spin asymmetries $A^{W^{\pm}}_L$ in Drell-Yan type processes.
Moreover, the model result of the quantity $\Delta \Sigma$~(which is usually stated as the quark spin content of the proton) just falls on the parametrized results. Therefore the quark-antiquark asymmetry~[i.e., $\Delta q_s(x) \neq \Delta \bar q_s(x)$] plays an important role for a comprehensive understanding of the nucleon spin structure.

\section*{Acknowledgements}

 This work is supported by National Natural Science Foundation of China (Grant No.~11475006).


\end{document}